\documentclass[pre,preprint,showpacs,amsmath]{revtex4}

\usepackage{graphicx}
\usepackage{bm}% bold math

\begin{document}

\title{Construction of the discrete breathers and a simple physical interpretation of their existence}
\author{G.M. Chechin}
  \email{chechin@aaanet.ru}
\author{G.S.Dzhelauhova}
\affiliation{South Federal University (Former Rostov State
University), Russia}
\date{\today}
\begin{abstract}
We present a simple numerical method for the discrete breather
construction based on the idea of the pair synchronization of the
particles involved in the breather vibration. It can be used for
obtaining exact breather solutions in nonlinear Hamiltonian lattices
of different types. We illustrate the above method using chains of
the coupled Duffing oscillators. With some additional approximation,
the pair synchronization method leads to a very simple physical
interpretation of the existence of the exact breathers as strictly
time-periodic and spatially localized dynamical objects.
\end{abstract}
\pacs{05.45.Yv, 63.20.Pw, 63.20.Ry} \maketitle

\section{Introduction}

The concept of the discrete breathers or intrinsic localized modes
in nonlinear Hamiltonian lattices have been proposed by Sievers and
Takeno two decades earlier in~\cite{S-T}. Since that time, there
appeared a vast number of papers considering different physical,
mathematical and numerical aspects of the discrete breathers
dynamics. The appropriate references can be found in many review
papers (see, for example, \cite{Aubry1, Aubry2, Flach1, Flach2}).

Discrete breathers represent spatially localized and time-periodic
excitations in nonlinear lattices. As we have already discussed
in~\cite{Chechin}, the possibility of the existence of such
dynamical objects is not obvious. Indeed, because of spatial
localization, different particles of the lattices vibrate with
essentially different amplitudes. On the other hand, it is typical
for nonlinear systems that frequencies depend on the amplitudes of
vibrating particles. Therefore, one may ask: "How can a discrete
breather exist as a \emph{strictly time-periodic} dynamical object
despite of its spatial localization"? In the present paper, we will
try to give an explicit answer to this question.

There exist several rigorous existence proofs for breathers in
networks of weakly coupled anharmonic oscillators reviewed, for
example, in~\cite{Aubry2}. The first existence proof of the discrete
breathers was obtained by the principle of anticontinuity
in~\cite{Mackay-Aubry}. The authors have demonstrated this approach
for the chain of linear coupled nonlinear oscillators where
anticontinual limit corresponds to zero coupling. The main idea of
this approach is connected with application of the implicit function
theorem in the space of time-periodic solutions at the discrete
breather period.

There exist a number of methods for breathers construction at
computer accuracy provided that their frequency (and its multiples)
is far enough from resonance with linear phonon spectrum
\cite{Aubry2, Flach2}. As was emphasized by Aubry in \cite{Aubry2},
the Newton method is used practically in all such methods and "it is
nothing but a numerical implementation of the implicit function
theorem" used for rigorous proof of the DBs existence in
\cite{Mackay-Aubry}. As a certain variation, in some numerical
procedures, the Newton method is used for finding the Fourier
coefficients of the exact breathers solutions \cite{Flach-94,
Flach2}.

Unfortunately, the above methods, as well as the rigorous existence
proofs, do not lead (at least, from our point of view) to any simple
physical interpretation of the DBs existence \footnote{Nevertheless,
let us note that a certain physical interpretation of the DBs
existence can be found in~\cite{Ovch-1969}. We will comment on this
work in Sec.~\ref{sec4} of the present paper.}).

In the present paper, we develop a new simple method for breather
construction (Sec.~\ref{syn2}) which is able to produce breathers
solutions with high precision (Sec.~\ref{sec3}) and, on the other
hand, leads to an explicit physical interpretation of DBs existence
(Sec.~\ref{sec4}).

\section{Method of pair synchronization \label{syn2}}

Below we present the method of pair synchronization (PS-method) for
discrete breather construction in nonlinear Hamiltonian lattices.
Despite of the applicability of this method for wide class of such
lattices, we prefer to demonstrate it with a simple case of the
monoatomic chain of the linear coupled Duffing oscillators for
clarity of exposition \footnote{This model belongs to the class of
mechanical systems treated in~\cite{Mackay-Aubry} where the first
rigorous proof of the discrete breathers existence was presented.}.
The generalization of the PS-method for other lattices seems to be
more or less straightforward.

The term "method of pair synchronization" is connected with the
specific nature of its algorithm. Indeed, the PS-method represents
an iterative procedure which implies subsequent synchronization of
the vibrations of each pair of the breather's particles.

\begin{figure}
\includegraphics{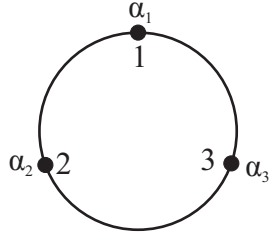}
\caption{\label{fig1} Scheme of the initial conditions for symmetric
breather in the chain~(\ref{eq1}) with $N=3$ particles and periodic
boundary conditions: $x_1(0)=x_2(0)=x_3(0)=0$; \
$\dot{x}_1(0)=\alpha_1$, \ $\dot{x}_2(0)=\alpha_2$,
$\dot{x}_3(0)=\alpha_3$.}
\end{figure}

The dynamical equations for the chain of the hard Duffing
oscillators with linear coupling can be written as follows
\begin{equation}\label{eq1}
\ddot{x}_i+x_i+x_i^3=\gamma (x_{i+1}-2x_i+x_{i-1}), \ \ \ i=1..N.
\end{equation}
Introducing the \emph{periodic} boundary conditions
\begin{equation}\label{eq2}
x_{N+1}(t)=x_1(t), \ \ \ x_0(t)=x_N(t),
\end{equation}
we will consider the simple symmetric breathers in the chains (1)
with $N=3, 5, 7, 9... etc$. particles. Because of the periodic
boundary conditions~(\ref{eq2}), we can imagine that our chain
represents a \emph{ring} depicted in its equilibrium state in Figs.
\ref{fig1}, \ref{fig5}, \ref{F50}.

\begin{figure}[h!t!b!]
%\centering
\includegraphics[width=100mm,height=55mm]{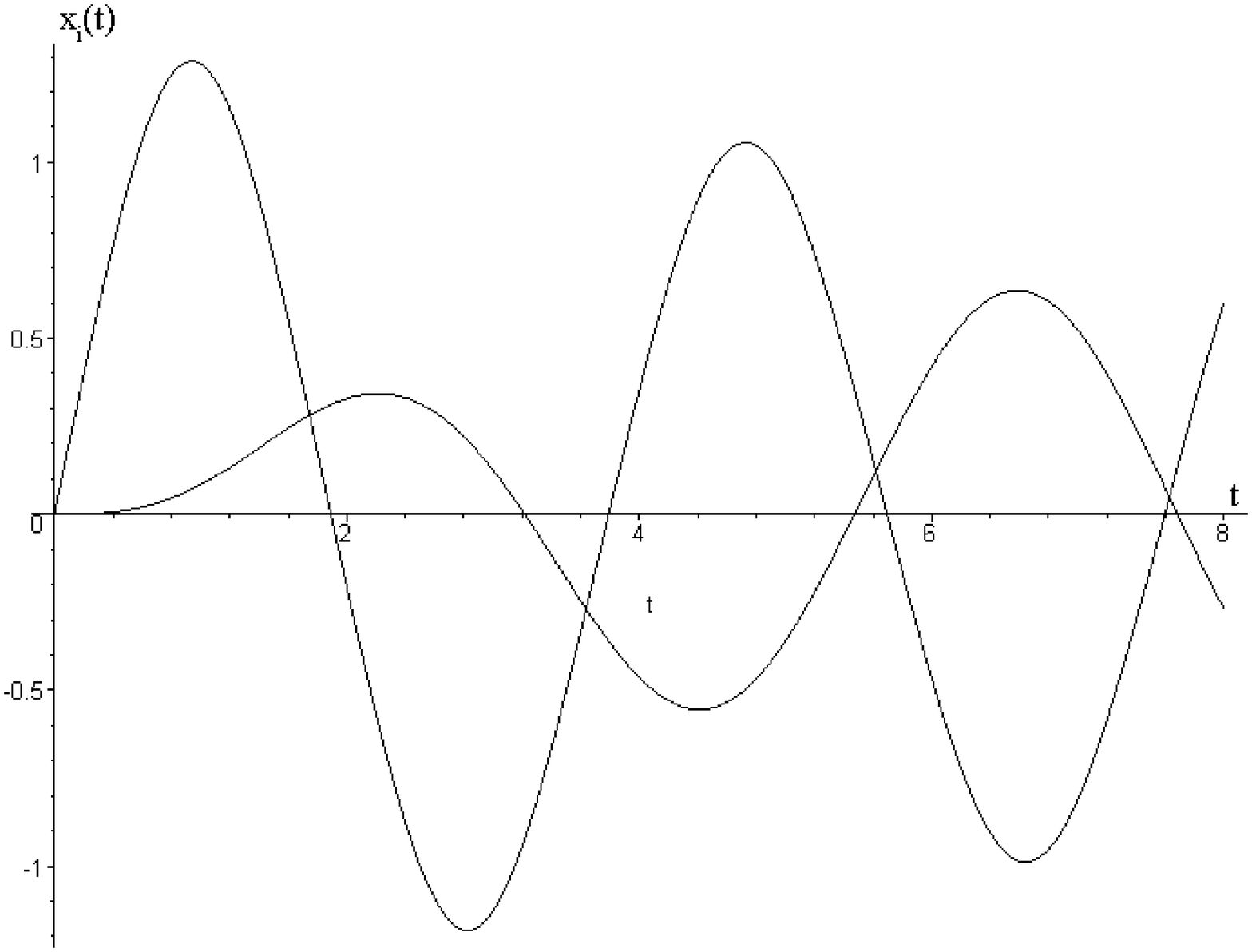}
\caption{\label{F1} Start of the pair synchronization method for the
chain with $N=3$ particles [$\dot{x}_1(0)=2$, \
$\dot{x}_2(0)=\dot{x}_3(0)=0$].}
\end{figure}

\begin{figure}[h!t!b!]
%\centering
\includegraphics[width=100mm,height=55mm]{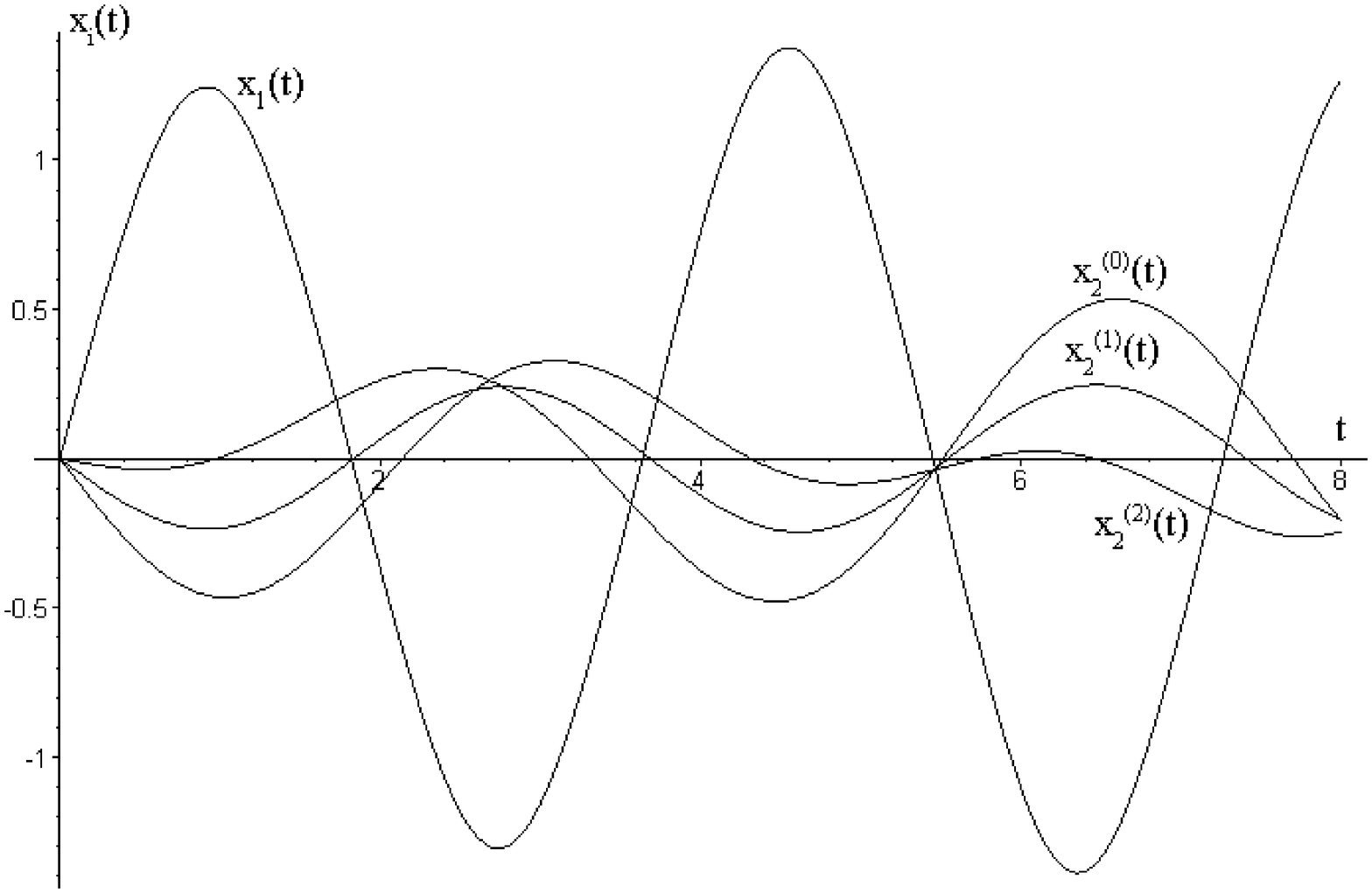}
\caption{\label{F2} The functions $x_1(t)$ and $x_2(t)$ for
different values of $\dot{x}_2(0)=\alpha_2$ for the
model~(\ref{eq1}) with $\gamma=0.3$ [$\alpha_2^{(0)}=-0.7, \
\alpha_2^{(1)}=-0.4, \ \alpha_2^{(2)}=-0.1$].}
\end{figure}

\begin{figure}[h!t!b!]
%\centering
\includegraphics[width=100mm,height=55mm]{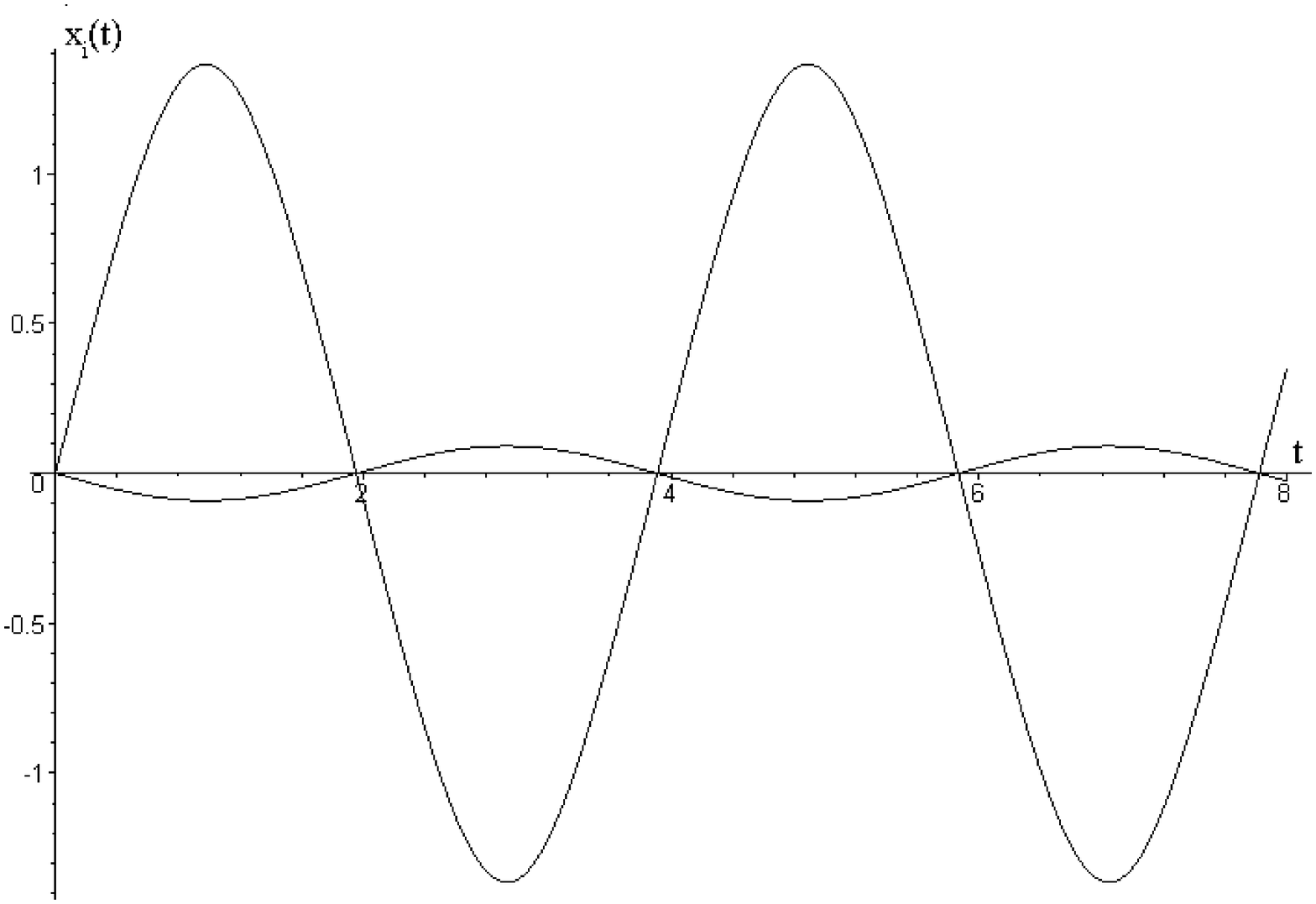}
\caption{\label{F3} Duffing chain (\ref{eq1}) with $N=3$ and
$\gamma=0.1$: the final result of the synchronization of the
vibrations of the particles ($\alpha_1=2$, \ $\alpha_2=-0.144656$).}
\end{figure}

We start with the case $N=3$ (see Fig.~\ref{fig1}). Let the first
particle be the central particle of the breather and, therefore, its
amplitude is greater than amplitudes of the peripheral particles,
i.e.
\begin{equation}\label{eq3}
|x_1(t)|>|x_2(t)|, |x_3(t)|.
\end{equation}
Moreover, the relation $x_2(t)=x_3(t)$ must hold as far as we search
for the symmetric breather.

To construct such breather, the following initial conditions can be
used for integrating the dynamical equations~(\ref{eq1}):
\begin{equation}\label{eq4}
x_1(0)=x_2(0)=x_3(0)=0, \ \dot{x}_1(0)=\alpha_1, \
\dot{x}_2(0)=\dot{x}_3(0)=\alpha_2.
\end{equation}
Thus, we suppose that all particles of the chain are in their
equilibrium positions at the initial instant, but they possess
certain velocities which we denote by $\alpha_i$ \ ($i=1..N$).

There are two parameters from which our breathers depend on, namely,
$\gamma$ entering the dynamical equations~(\ref{eq1}) and the
initial velocity $\alpha_1$ of the central breather's particle.
Choosing some reasonable values of these parameters and supposing
$\alpha_2=0$, we integrate equations~(\ref{eq1}) for $N=3$ over time
interval containing, at least, one zero of the function $x_1(t)$ and
one zero of $x_2(t)$ \ (see Fig.~\ref{F1}). Now let us \emph{vary}
the initial velocity $\dot{x_2}(0)=\alpha_2$ of the second particles
aiming at \emph{coincidence} of the first zero of $x_2(t)$ with that
of $x_1(t)$. In other words, we try to \emph{synchronize} the
oscillations of the first particle [$x_1(t)$] with those of two
neighboring particles [$x_2(t)\equiv x_3(t)$].

Different numerical methods can be applied for such purpose, for
example, the shooting method using the idea  of dichotomy. Zeros of
the functions $x_1(t)$ and $x_2(t)$ can be found by the
Newton-Raphson method or by some other well-known numerical methods.

Some comments are appropriate at this stage.

Let $i_1=1, \ 2, \ 3,...$ labels the subsequent zeros of the
function $x_1(t)$ and $i_2=1, \ 2, \ 3,...$ labels those of the
function $x_2(t)$. In principle, we can superpose zeros of the
functions $x_1(t)$ and $x_2(t)$ with \emph{different} values $i_1$
and $i_2$. As a consequence, we obtain discrete breathers of
\emph{different} types (see Sec.~\ref{sec3}).

In this section, we consider only the \emph{single-frequency}
breathers, i.e. the breathers for which all particles of the chain
oscillate with one and the same frequency  $\omega_j=\omega_b$ \
($j=1..N$). This is not a general case. Indeed, one can imagine
\emph{two-frequency}, \emph{three-frequency}, etc. breathers whose
particles possess \emph{different}, but multiple frequencies (for
example, if $\omega_1=2\omega_2$ \ or \ $\omega_1=3\omega_2$ for the
chain with $N=3$ particles, we have two-frequency breather). In such
a case, the breather frequency $\omega_b$ is equal to the
\emph{minimal} frequency among all possible $\omega_j$ \ ($j=1..N$)
of the individual particles. Therefore, all the particles vibrate
with the frequencies which prove to be multiples of $\omega_b$.

In order to avoid some misunderstanding, let us note that the
single-frequency breather is \emph{not monochromatic}, i.e. although
all the breather's particles possess identical vibrational
frequencies, the Fourier series of periodic functions $x_i(t)$
contains not one, but a certain set of different harmonics.

Some comments on the terminology are appropriate at this point. The
terms single-frequency and many-frequency breathers, which we use in
the present paper, disagree with those used in~\cite{Flach-94}.
Indeed, in the latter paper, a single-frequency discrete breather
means the conventional strictly time-periodic dynamical object,
while a many-frequency breather discribes a \emph{quasiperiodic}
motion on an $n$-dimensional torus ($n$ basic frequencies correspond
to its Fourier decomposition). In contrast, considering only
strictly \emph{time periodic} breathers, we divide them into
\emph{different classes}: in single-frequency breathers, all the
particles vibrate with one and the same frequency, while in
many-frequency breathers, different particles possess
\emph{different}, but divisible frequencies (see below for more
details).

Let us return to the discussion of the numerical procedure of the
pair synchronization for the chain with $N=3$ particles.

In Fig.~\ref{F2} we depict the function $x_2(t)$ for several values
$\alpha_2$ of the initial velocity of the second particle. From this
figure, one can see how the position of the first zero of $x_2(t)$
changes with changing $\dot{x}_2(0)=\alpha_2$. Note that changing of
$\alpha_2$ leads not only to a shift of the zero of $x_2(t)$, but
also to a certain shift of $x_1(t)$, but the latter turns out to be
rather small \footnote{This is a consequence of the relations
$\gamma\ll1$, \ $|\dot{x_2}(0)|\ll |\dot{x_1}(0)|$.} and is not
noticeable in Fig.~\ref{F2}.

It is obvious from Fig.~\ref{F2}, that $\alpha_2^{(1)}$ and
$\alpha_2^{(2)}$ provide a "fork" to which we can apply the method
of dichotomy to attain concidence of the zeros of the functions
$x_1(t)$ and $x_2(t)$ with a chosen level of accuracy. The final
result of the discussed synchronization is shown in Fig.~\ref{F3},
where one can see a single-frequency breather for the chain with
$N=3$ particles.

The synchronization of the vibrations of the first and second
particles we call $S(1,2)$ procedure. Complete realization of this
procedure for the case of the single-frequency breather leads to the
full coincidence of all the zeros of the functions $x_1(t)$ and
$x_2(t)$ with equal labels $i_1$ and $i_2$. In general case, the
similar procedure of synchronization of the vibrations of $i$-th and
$j$-th particles will be called $S(i, j)$ procedure.

Now, let us construct a single-frequency symmetric breather for the
chain~(\ref{eq1}) with $N=5$ particles (see Fig.~\ref{fig5}).
Choosing resonable values of $\gamma$ and $\alpha_1$, we have to
find such values of $\alpha_2$ and $\alpha_3$ which lead to the
equal vibrational frequencies of all the particles of our chain. In
other words, we want to synchronize vibrations of the central
particle [$x_1(t)$] with those of peripheral particles
[$x_2(t)=x_5(t)$ and $x_3(t)=x_4(t)$].

\begin{figure}
\includegraphics{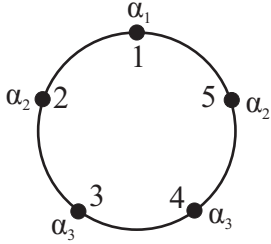}
\caption{\label{fig5} Scheme of the initial conditions for the
five-particles chain.}
\end{figure}

Let us apply the method of pair synchronization for discrete
breather construction in the considered chain. The algorithm of the
method can be described as follows. We synchronize vibrations of the
particles 1 and 2 with aid of the above described procedure $S(1,
2)$\footnote{Note that by performing this procedure we do not worry
about vibrational frequencies of all the other particles.}. Then we
synchronize vibrations of the particles 2 and 3 with the aid of the
procedure $S(2, 3)$. The sequence of these two procedures, $S(1,
2)$, \ $S(2, 3)$, forms one \emph{iterative cycle} of the pair
synchronization method. We must repeat such cycles, namely, $S(1,
2); \ S(2,3)| \ S(1, 2); \ S(2,3)| \ S(1, 2); \ S(2,3)|...$, \ up to
attain a desirable level of accuracy. Indeed, the coincidence of
zeros of $x_1(t)$ and $x_2(t)$, after finishing the procedure
$S(1,2)$, will be slightly disturbed because of applying the
procedure $S(2,3)$ which changes $\alpha_2$, \ $\alpha_3$, and we
must recorrect the accuracy of this coincidence by the next
iterative cycle. One can hope for convergence of such iterative
process, at least, for the case of sufficiently strong space
localization of the discrete breather.

The generalization of the considered method to a chain with
arbitrary $N$ is straightforward: we repeat iterative cycles
\begin{equation}\label{eq20}
\{S(1,2); \ S(2, 3); \ S(3, 4);... \ S(N-1, N)\}
\end{equation}
up to obtaining the discrete breather with resonable precision. Note
that the method of pair synchronization can be regarded as a version
of the relaxation procedure used in many different numerical
methods.

Note that considering chains with subsequently increasing number of
particles ($N=3, \ 5, \ 7, \ 9, \ 11,...$), we can use the final set
of initial velocities ($\alpha_1$, \ $\alpha_2$, \ $\alpha_3$,... )
of the chain with $N=m$ in order to begin calculations for the chain
with $N=m+2$, assuming two last velocities equal to zero.

\section{Some numerical results\label{sec3}}

Below, we present some results of the discrete breathers
construction with the aid of the pair synchronization method for a
number of mechanical models and for different kinds of discrete
breathers. The application of this method to other oscillatory
chains and multidimensional lattices will be discussed in future
publications.

\subsection{Linear coupled Duffing oscillatory chain}

Let us consider the model~(\ref{eq1}) for different values of the
coupling parameter $\gamma$ and different values of the initial
velocity $\alpha_1$ of the central breather's particle.

\begin{figure}[h!t!b!]
%\centering
\includegraphics[width=71mm,height=55mm]{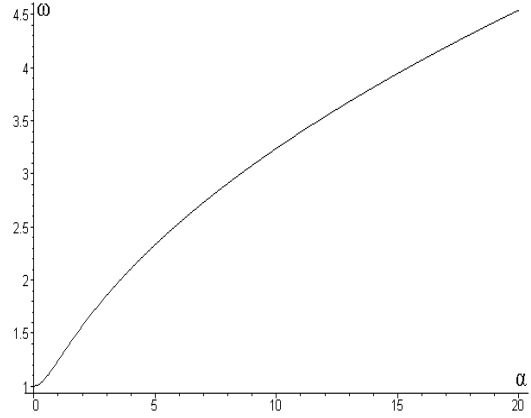}
\caption{\label{F10} The function $\omega(\alpha)$ determining
dependence of the Duffing oscillator frequency $\omega$ on the
initial velocity $\alpha$.}
\end{figure}

The appropriate initial velocities $\dot{x}_i(0)=\alpha_i$, \
$i=2..N$ are necessary for a start of the pair synchronization
procedure. For $\gamma\lesssim 1$ one can obtain these velocities
from the approximate formula, which is derived in Sec.~\ref{sec4}:
\begin{equation}\label{eq50}
\alpha_{i+1}=-\frac{\gamma}{\omega_b^2-\omega_0^2}\alpha_i, \ \ \
(i=1..N-1).
\end{equation}
Here $\omega_b$ is the breather frequency, which can be roughly
approximated as $\omega_b=\omega(\alpha_1)$ with $\omega(\alpha_1)$
being the frequency of a single Duffing oscillator, while
$\omega_0^2=1+2\gamma$. The function $\omega(\alpha)$ is determined
by the solution of the Duffing equation $\ddot{x}+x+x^3=0$ and we
depict it in Fig.~\ref{F10} for obtaining rough estimates of the
breather frequency $\omega_b$.

In Table~\ref{t1}, we present the velocity and amplitude breather
profiles, as well as the breather frequency $\omega_b$, for the
chains with different number of particles $N=3$, $5$, $7$, $9$,
$11$, $13$ for $\gamma=0.3$, \ $\alpha_1=2$. In the first column of
each fragment of this table the indices ($n$) of the particles are
given (for the central breather particle $n=0$). In the following
columns, we present the initial velocity ($\alpha_n$) and
vibrational amplitude ($A_n$) of each particle of the considered
chain.
\begin{table}
  \centering
  \caption{\label{t1} Discrete breathers for the Duffing chain (\ref{eq1}) for $\gamma=0.3$ and different number of particles ($N$).}
\begin{tabular}{|c|c|c|c|c|c|c|c|c|}
  \hline
  \multicolumn{3}{|c|}{$N=3$; $\omega_b=1.70714$} & \multicolumn{3}{|c|}{$N=5$; $\omega_b=1.70670$} & \multicolumn{3}{|c|}{$N=7$; $\omega_b=1.70711$} \\
  \hline
  n & $\alpha_n$ & $A_n$ & n & $\alpha_n$ & $A_n$ & n & $\alpha_n$ & $A_n$ \\
  \hline
      &           &         &   &          &            & -3& -0.024052  & 0.014091 \\
      &           &         & -2& 0.099350 & 0.058247   & -2& 0.129353   & 0.075822 \\
   -1 & -0.406807 & 0.240930& -1& -0.530612& 0.313171   & -1& -0.538053  & 0.317510 \\
   0  & 2         & 1.261441& 0 & 2        & 1.254590   & 0 & 2          & 1.254181 \\
   1  & -0.406807 & 0.240930& 1 & -0.530612& 0.313171   & 1 & -0.538053  & 0.317510 \\
      &           &         & 2 & 0.099350 & 0.058247   & 2 & 0.129353   & 0.075822 \\
      &           &         &   &          &            & 3 & -0.024052  & 0.014091 \\
  \hline
  \multicolumn{3}{|c|}{$N=9$; $\omega_b=1.70714$} & \multicolumn{3}{|c|}{$N=11$; $\omega_b=1.70714$} & \multicolumn{3}{|c|}{$N=13$; $\omega_b=1.70715$} \\
  \hline
  n & $\alpha_n$ & $A_n$ & n & $\alpha_n$ & $A_n$ & n & $\alpha_n$ & $A_n$ \\
  \hline
      &            &            &   &              &            & -6 & 0.000339   & 0.000197  \\
      &            &            & -5& -0.001403    & 0.000822& -5 & -0.001825   & 0.001069  \\
   -4 & 0.005812   & 0.003404   & -4& 0.007555     & 0.004426 & -3 & 0.007657   & 0.004485  \\
   -3 & -0.031273  & 0.018319   & -3& -0.031694    & 0.018566  & -3 & -0.031718  & 0.018581  \\
   -2 & 0.131113   & 0.076849   & -2& 0.131215     & 0.076909  & -2 & 0.131221      & 0.076913  \\
   -1 & -0.538494  & 0.317761   & -1& -0.538519    & 0.317776   & -1 & -0.538520     & 0.317777  \\
   0  &  2         & 1.254159   & 0 & 2            & 1.254160   & 0 & 2             & 1.254160 \\
   1  & -0.538494  & 0.317761   & 1 & -0.538519    & 0.317776   & 1 & -0.538520      & 0.317777   \\
   2  & 0.131113   & 0.076849   & 2 & 0.131215     & 0.076909  & 2 & 0.131221       & 0.076913  \\
   3  & -0.031273  & 0.018319   & 3 & -0.031694    & 0.018566  & 3 & -0.031718   & 0.018581  \\
   4  & 0.00581    & 0.003404   & 4 & 0.007555     & 0.004426 & 4 & 0.007657    & 0.004485  \\
      &            &            & 5 & -0.001403    & 0.000822& 5 & -0.001825   & 0.001069  \\
      &            &            &   &              &            & 6 & 0.000339    & 0.000197 \\
  \hline
\end{tabular}
\end{table}

From this table, one can see how the discrete breather for the
\emph{infinite} chain acquires its shape step-by-step with
increasing $N$. We restrict our consideration to  $N=13$ because, in
the case $\gamma=0.3$, the amplitudes of the particles which are the
most distant from the breather's center do not exceed $10^{-4}$ of
the amplitude of the central particle.

In Table~\ref{t2}, we present information similar to that in
Table~\ref{t1} for the chain with $N=9$ particles, but for different
values of the coupling parameter $\gamma$ \
$(0.3\leq\gamma\leq1.5)$. From this table, one can see that a degree
of localization becomes worser with increasing $\gamma$.
\begin{table}
  \centering
  \caption{\label{t2} Discrete breathers for the Duffing chain (\ref{eq1}) for different values of the coupling parameter $\gamma$ ($N=9$).}
\begin{tabular}{|c|c|c|c|c|c|c|c|c|}
  \hline
  \multicolumn{3}{|c|}{$\gamma=0.3$; $\omega_b=1.70714$} & \multicolumn{3}{|c|}{$\gamma=0.5$; $\omega_b=1.85545$} & \multicolumn{3}{|c|}{$\gamma=0.7$; $\omega_b=2.03582$}\\
  \hline
  n & $\alpha_n$ & $A_n$ & n & $\alpha_n$ & $A_n$ & n & $\alpha_n$ & $A_n$ \\
  \hline
   -4 & 0.005812   & 0.003404 & -4 & 0.065523  & 0.022544 & -4& 0.100398  & 0.049322 \\
   -3 & -0.031273  & 0.018319 & -3 & -0.172127 & 0.087583 & -3& -0.350110 & 0.172155 \\
   -2 & 0.131113   & 0.076849 & -2 &  0.428520 & 0.229276 & -2& 0.759077  & 0.374592 \\
   -1 &-0.538494   & 0.317761 & -1 & -1.022931 & 0.557273 & -1& -1.417113 & 0.706940 \\
    0 & 2          & 1.254159 & 0  & 2         & 1.128836 & 0 & 2         & 1.013062 \\
    1 & -0.538494  & 0.317761 & 1  & -1.022931 & 0.557273 & 1 & -1.417113 & 0.706940 \\
    2 & 0.131113   & 0.076849 & 2  & 0.428520  & 0.229276 & 2 & 0.759077  & 0.374592 \\
    3 & -0.031273  & 0.018319 & 3  & -0.172127 & 0.087583 & 3 & -0.350110 & 0.172155 \\
    4 & 0.005812   & 0.003404 & 4  & 0.065523  & 0.022544 & 4 & 0.100398  & 0.049322 \\
  \hline
  \multicolumn{3}{|c|}{$\gamma=1$; $\omega_b=2.29177$} & \multicolumn{3}{|c|}{$\gamma=1.2$; $\omega_b=2.44835$} & \multicolumn{3}{|c|}{$\gamma=1.5$; $\omega_b=2.66735$} \\
  \hline
  n & $\alpha_n$ & $A_n$ & n & $\alpha_n$ & $A_n$ & n & $\alpha_n$ & $A_n$ \\
  \hline
     -4& 0.175463   & 0.076581  & -4 & 0.212225  & 0.086704 & -4 & 0.250808  & 0.094053   \\
     -3& -0.569282  & 0.248818  & -3 & -0.669348 & 0.273867 & -3 & -0.769750 & 0.289048 \\
     -2& 1.077299   & 0.472699  & -2 & 1.200646  & 0.492969 & -2 & 1.312950  & 0.494406  \\
     -1& -1.669057  & 0.737848  & -1 & -1.738912 & 0.717933 & -1 & -1.793009 & 0.677654 \\
     0 & 2          & 0.889146  & 0  & 2         & 0.828581 & 0 &    2       & 0.757359  \\
     1 & -1.669057  & 0.737848  & 1  & -1.738912 & 0.717933 & 1 & -1.793009  & 0.677654 \\
     2 & 1.077299   & 0.472699  & 2  & 1.200646  & 0.492969 & 2 & 1.312950   & 0.494406  \\
     3 & -0.569282  & 0.248818  & 3  & -0.669348 & 0.273867 & 3 & -0.769750  & 0.289048 \\
     4 & 0.175463   & 0.076581  & 4  & 0.212225  & 0.086704 & 4 & 0.250808   &  0.094053 \\
  \hline
\end{tabular}
\end{table}

\begin{figure}
\includegraphics[width=71mm,height=56mm]{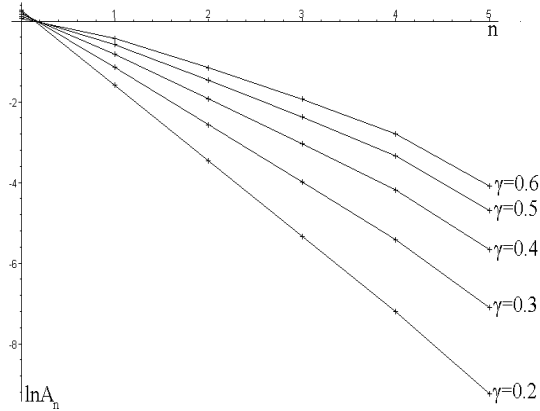}
\caption{\label{F15} The amplitudes $A_n$ for the symmetric
breathers in the Duffing chain~(\ref{eq1}) for different values
$\gamma$ (in the logarithmic scale).}
\end{figure}

Because the amplitudes $A_n$ of the particles decrease rapidly with
increasing their distance from the breather center, we depict them
 in the logarithmic scale for different $\gamma$ in
Fig.~\ref{F15}. From this picture one can see the \emph{exponential
decay }of the amplitudes of the peripheral particles. The degree of
the breather space localization decreases with increasing $\gamma$.

Let us note that the pair synchronization method can be often used
not only for sufficiently small coupling between the oscillators of
the lattice. From Table~\ref{t2}, we see that this method also
works,  in the case of the model~(\ref{eq1}), for $\gamma\sim1$.
Moreover, the pair synchronization method can be used for $\gamma>1$
and even for $\gamma\gg1$ (in such cases formula (\ref{eq50}) cannot
be used for obtaining an initial velocity profile). As an example,
let us give the set of the initial velocities \
$\dot{x}_i(0)=\alpha_i$ \ obtained by this method for the
single-frequency discrete breather in the chain~(\ref{eq1}) with
$N=7$ particles for $\gamma=20$: \ $\{\alpha_1=2, \
\alpha_2=-1.801427, \ \alpha_3=1.245828, \ \alpha_4=-0.444403\}$.
Existence and stability of the discrete breathers in chains with
large coupling will be discussed elsewhere.

\subsection{Many-frequency discrete breathers}

First of all let us comment on one point of the procedure of pair
synchronization. Considering this procedure in Sec.~\ref{syn2}, we
have aimed at coincidence of those zeros of two functions, $x_i(t)$
and $x_j(t)$, which are most close to the initial instant $t=0$.
This is not a general case. Indeed, one can imagine that $x_i(t)$
and $x_j(t)$ possess \emph{equal} periods $T_i=T_j=T$, but that they
do not go to zero \emph{simultaneously} at any \emph{internal}
instant $0<t<T$. Then we should aim at coincidence of the
\emph{second} zeros of the considered functions, i.e. the relation
$\dot{x}_i(T)=\dot{x}_j(T)=0$ must hold.

Now let us focus on constructing \emph{many-frequency} discrete
breathers. Let us remember that we coin this term for such spatially
localized dynamical objects whose particles vibrate with several
\emph{different} but \emph{divisible} frequencies. This means that
many-frequency breather proves to be a \emph{time-periodic} object
whose period $T$ is the largest of the multiple periods $T_1$, \
$T_2$, ..., $T_N$ of all the particles of the chain.

\begin{figure}
\includegraphics[width=100mm,height=55mm]{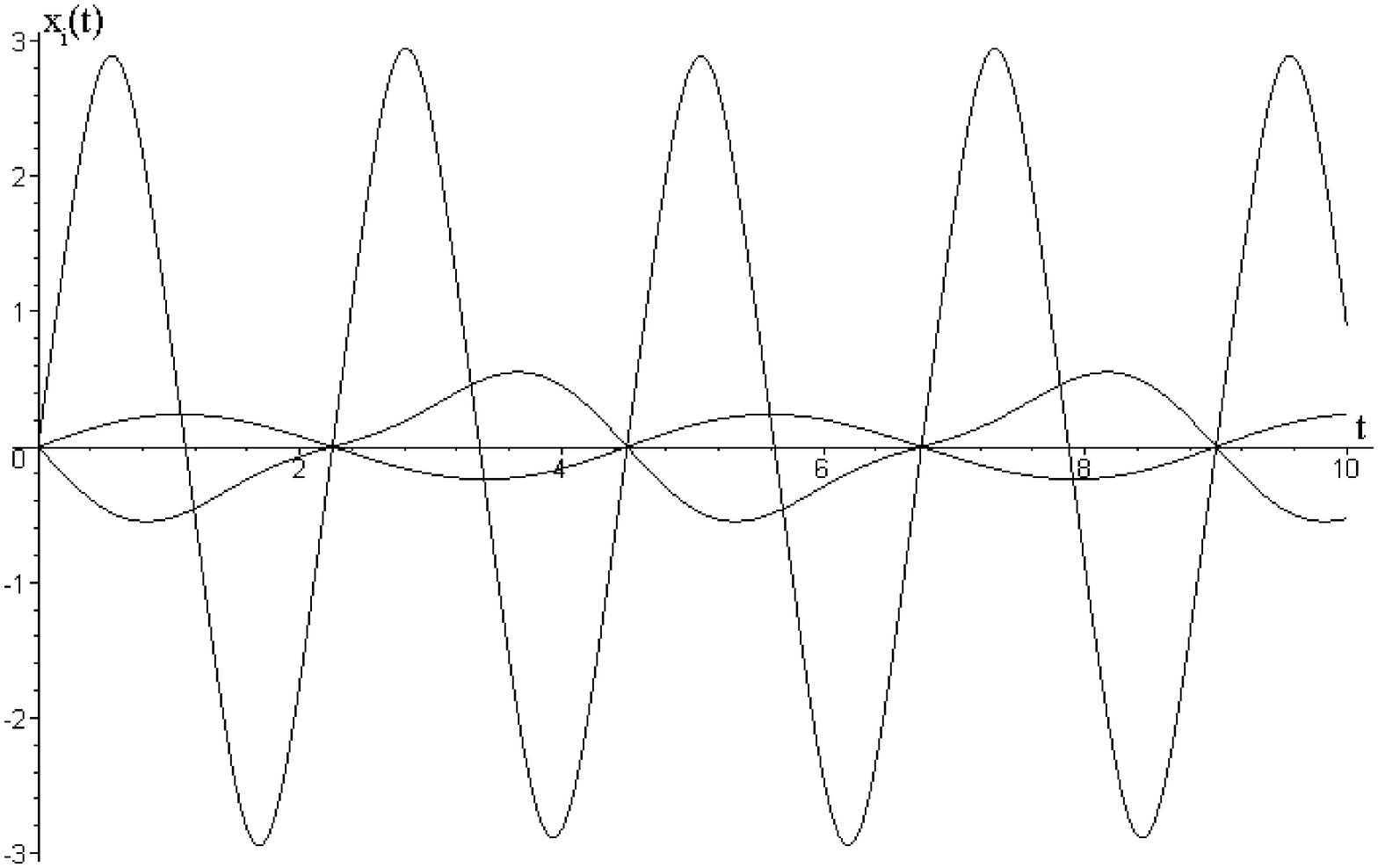}
\caption{\label{F16} Two-frequency symmetric breather in the Duffing
chain~(\ref{eq1}) for $\gamma=0.3$, \ $\alpha_1=7$, \ $N=5$.}
\end{figure}

In Fig.~\ref{F16}, we depict a \emph{two-frequency} symmetric
breather ($\omega_1=2\omega_2=2\omega_3$) in the  chain (\ref{eq1})
with $N=5$ particles for $\gamma=0.3$. Indeed, the first particle
$[x_1(t)]$ vibrates with the frequency twice larger than those of
all other particles $[x_2(t), \ x_3(t)]$. This breather with
$\omega_b=\omega_2(=\omega_3)=1.39615$ corresponds to the following
data: $\gamma=0.3$, \ $\alpha_1=7$, \ $\alpha_2=-1.064224$, \
$\alpha_3=0.352247$. Hereafter, we do not point out explicitly that
all the initial displacements are zero: $x_i(0)=0$, \ $i=1..N$.

\begin{figure}
\includegraphics[width=100mm,height=55mm]{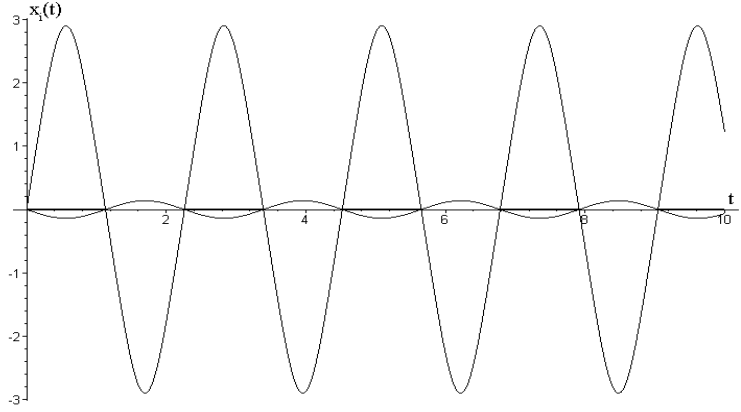}
\caption{\label{F17} Single-frequency symmetric breather in the
Duffing chain~(\ref{eq1}) for $\gamma=0.3$, \ $\alpha_1=7$, \ $N=5$.
Vibrations of the third particle is not visible, because of its
small amplitude.}
\end{figure}

Note that for the same $\gamma$ and $\alpha_1$, we can find a
\emph{single-frequency} breather
($\omega_1=\omega_2=\omega_3=2.77828$), as well (see Fig.~\ref{F17}
which was obtained for $\gamma=0.3$, \ $\alpha_1=7$, \
$\alpha_2=-0.379206$, \ $\alpha_3=0.017876$, \ $N=5$).

\begin{figure}
\includegraphics[width=100mm,height=55mm]{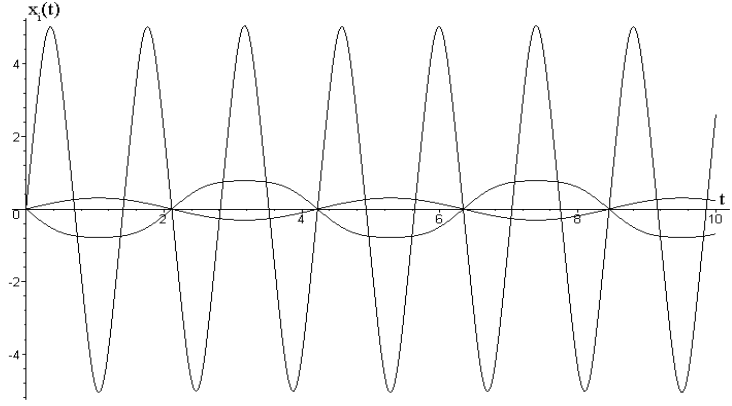}
\caption{\label{F18} Two-frequency symmetric breather in the Duffing
chain~(\ref{eq1}) for $\gamma=0.3$, \ $\alpha_1=19$, \ $N=5$ \
($\omega_1=3\omega_2=3\omega_3$). This breather is unstable.}
\end{figure}

In Fig.~\ref{F18}, we depict the \emph{two-frequency} breather with
$\omega_b=\omega_2=1.48666$ whose particles possess the following
frequencies: $\omega_1=3\omega_2=3\omega_3$. This breather
corresponds to the following initial data: $\gamma=0.3$, \
$\alpha_1=19$, \ $\alpha_2=-1.593445$, \ $\alpha_3=0.456080$, \
$N=5$.

\begin{figure}
\includegraphics[width=100mm,height=55mm]{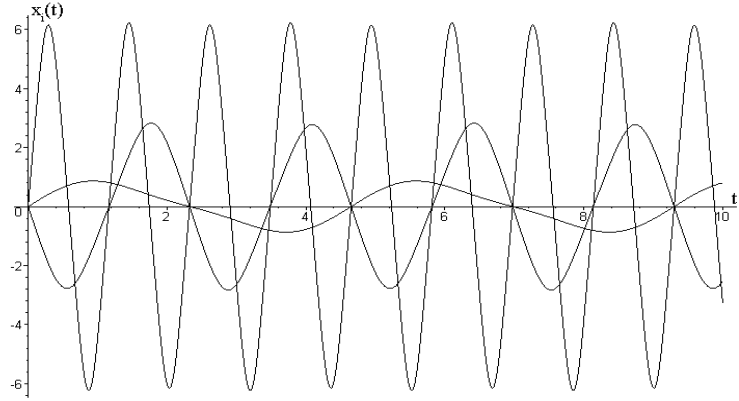}
\caption{\label{F19} Three-frequency symmetric breather in the
Duffing chain~(\ref{eq1}) for $\gamma=0.3$, \ $\alpha_1=28$, \ $N=5$
\ ($\omega_1=2\omega_2=4\omega_3$). This breather is unstable.}
\end{figure}

In Fig.~\ref{F19}, a \emph{three-frequency} breather with
$\omega_1=2\omega_2$, \ $\omega_2=2\omega_3$ is shown. It
corresponds to the following initial data: $\gamma=0.3$, \
$\alpha_1=28$, \ $\alpha_2=-7.018426$, \ $\alpha_3=1.472607$, \
$N=5$. The breather frequency is $\omega_b=\omega_3=1.35108$.

Note that the pair synchronization method providing us with  an
initial breather profile does not guarantee the breather stability.
For example, the breather depicted in Figs.~\ref{F18},~\ref{F19}
proves to be unstable dynamical object, and this fact can be checked
by straightforward integration of the dynamical
equation~(\ref{eq1}), as well as with the aid of the Floquet method.

\begin{figure}
\includegraphics[width=100mm,height=55mm]{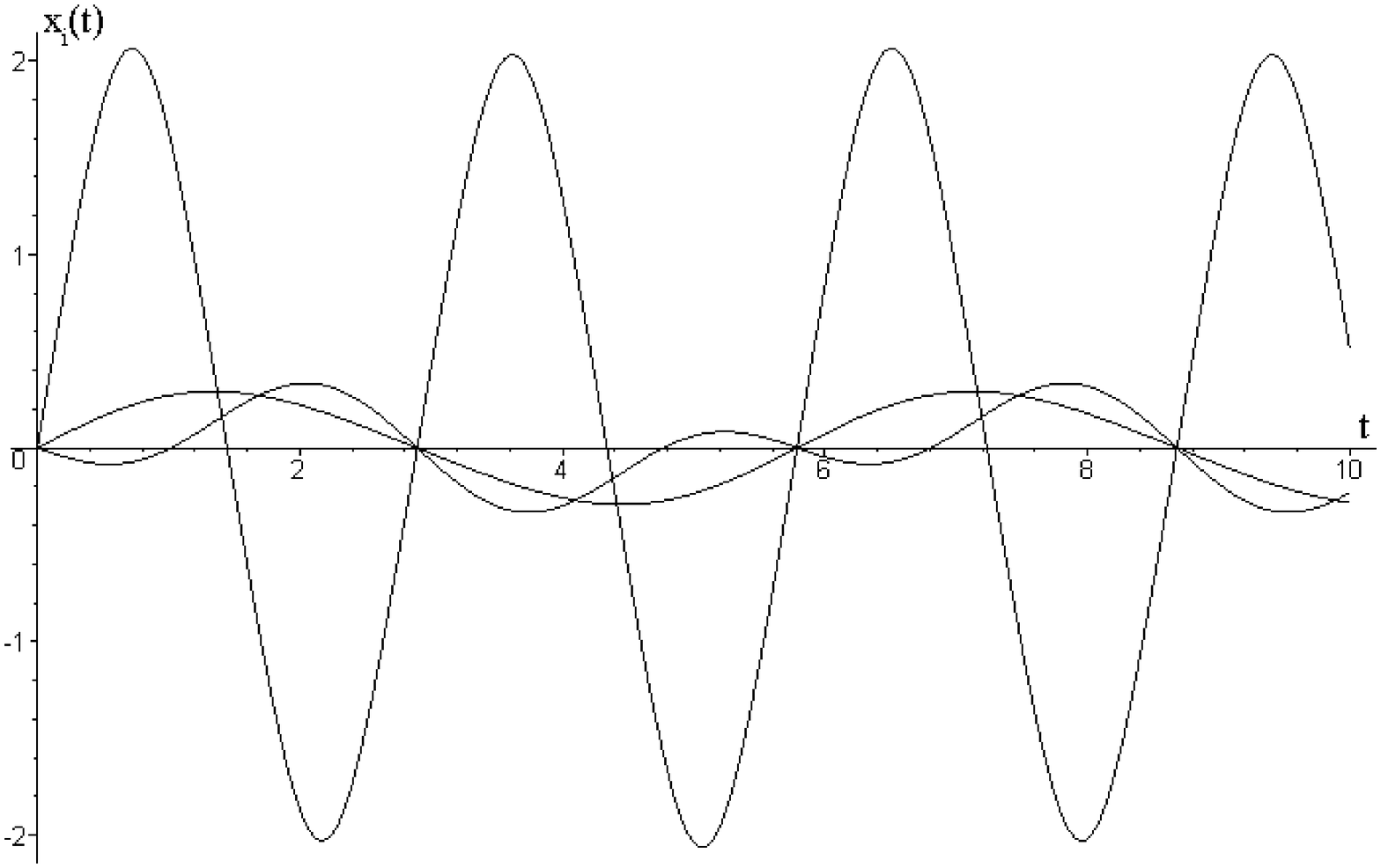}
\caption{\label{F20} A nontrivial two-frequency breather in the
Duffing chain~(\ref{eq1}) for $\gamma=0.3$, \ $\alpha_1=4$, \ $N=5$
\ ($\omega_1=2\omega_2=2\omega_3=2\omega_4$).}
\end{figure}

Finally, in Fig.~\ref{F20}, we demonstrate a nontrivial example of
two-frequency breathers ($\gamma=0.3, \alpha_1=4,
\alpha_2=-0.233618, \alpha_3=0.350414$, \ $N=5$). Indeed the
function $x_2(t)$, unlike $x_3(t)$, possesses two different maxima
(minima) within one breather  period which coincides with that of
$x_2(t)$ and $x_3(t)$. Note that vibrational amplitudes of the
second and third particles are of the same order of smallness.

Note that, for simplicity, in many examples of this section we
consider only chains with small number of particles ($N$). This is
not an essential restriction because, in every such case, the
synchronization procedure can be easily continued for much greater
values of $N$.

Let us emphasize once more that, unlike the terminology of the paper
\cite{Flach-94}, our\emph{ many-frequency} breathers \emph{are
time-periodic} dynamical objects, but different particles vibrate
with different (divisible) frequencies $\omega_i$ \ ($i=1..N$). We
would like to note that one can found in~\cite{Aubry1} a certain
mention of multibreathers whose particles vibrate with different but
commensurate frequencies. However, we do not know any papers with
detailed analysis of such dynamical objects and it seems that our
many-frequency breathers are not fully identical with the above
mentioned multibreathers.

\subsection{Constructing of antisymmetric breathers}

Up to this point, we have considered only the \emph{symmetric}
discrete breathers, i.e. breathers of the form $\{... x_3(t),
x_2(t), x_1(t), x_0(t), x_1(t), x_2(t), x_3(t),...\}$. Often they
 are called Sievers-Takeno modes.

\begin{figure}
\includegraphics[width=100mm,height=55mm]{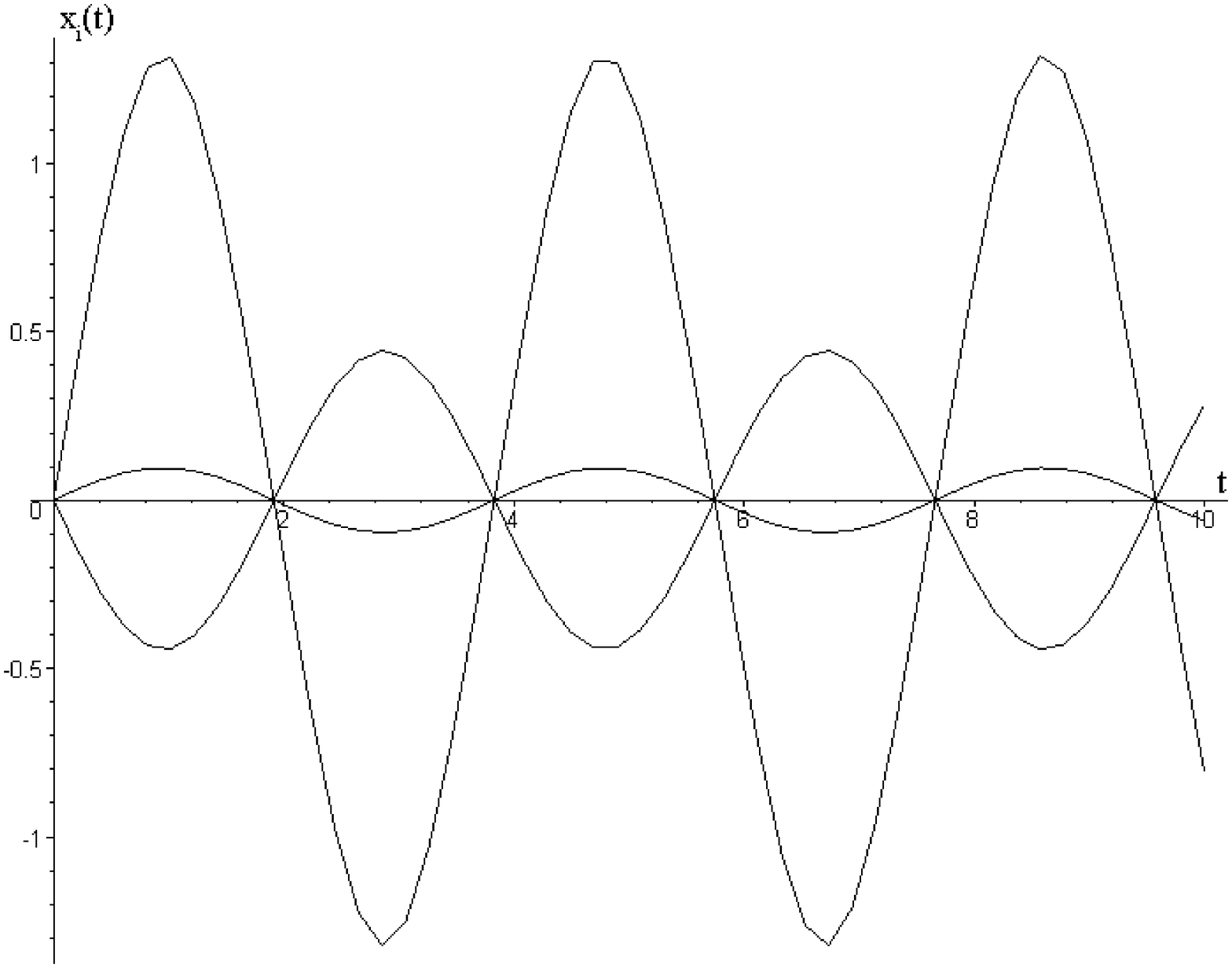}
\caption{\label{F30} Single-frequency antisymmetric breather in the
Duffing chain~(\ref{eq1}) for $N=6$, \ $\gamma=0.3$, \
$\alpha_1=2$.}
\end{figure}

The pair synchronization method can be also used for constructing
\emph{antisymmetric} breathers which possess the following form
$\{... -x_3(t), -x_2(t), -x_1(t), x_1(t), x_2(t), x_3(t),...\}$.
Such breathers are usually called Page modes. In Table~\ref{t10}, we
present initial velocities ($\alpha_i$) and amplitudes ($A_i$) of
all particles of the antisymmetric breather for the Duffing chain
(\ref{eq1}) with $N=8$ particles. One can see the time evolution of
this breather in Fig.~\ref{F30}.

\begin{table}
  \centering
  \caption{\label{t10} A single-frequency antisymmetric breather in the Duffing chain (\ref{eq1})  for different values of the coupling parameter $\gamma$ ($N=8$).}
\begin{tabular}{|c|c|c|c|c|c|c|c|c|}
  \hline
  \multicolumn{3}{|c|}{$\gamma=0.3$; $\omega_b=1.74894$} & \multicolumn{3}{|c|}{$\gamma=0.5$; $\omega_b=1.89609$} & \multicolumn{3}{|c|}{$\gamma=0.7$; $\omega_b=2.05935$}\\
  \hline
  n & $\alpha_n$ & $A_n$ & n & $\alpha_n$ & $A_n$ & n & $\alpha_n$ & $A_n$ \\
  \hline
   -4 & 0.026418   & 0.015106 & -4 & 0.137987  & 0.072790  & -4& 0.379916  & 0.184678 \\
   -3 & -0.102000  & 0.058347 & -3 & -0.300923 & 0.158896  & -3& -0.604583 & 0.294409 \\
   -2 & 0.466790   & 0.268490 & -2 & 0.806715  & 0.428828  & -2& 1.148316  & 0.563216 \\
   -1 & -2         & 1.215605 & -1 & -2        & 1.100366  & -1& -2        & 1.000219 \\
    1 & 2          & 1.215605 & 1  & 2         & 1.100366  & 1 & 2         & 1.000219 \\
    2 & -0.466790  & 0.268490 & 2  & -0.806715 & 0.428828  & 2 & -1.148316 & 0.563216 \\
    3 & 0.102000   & 0.058347 & 3  & 0.300923  & 0.158896  & 3 & 0.604583  & 0.294409 \\
    4 & -0.026418  & 0.015106 & 4  & -0.137987 & 0.072790  & 4 & -0.379916 & 0.184678 \\
  \hline
  \multicolumn{3}{|c|}{$\gamma=1$; $\omega_b=2.31566$} & \multicolumn{3}{|c|}{$\gamma=1.2$; $\omega_b=2.48400$} &  \multicolumn{3}{|c|}{$\gamma=1.4$; $\omega_b=2.64741$}\\
  \hline
  n & $\alpha_n$ & $A_n$ & n & $\alpha_n$ & $A_n$ &  n & $\alpha_n$ & $A_n$  \\
  \hline
   -4 & 0.939477  & 0.407302 & -4 & 1.388243   & 0.562493 & -4 & 1.873498  & 0.714152 \\
   -3 & -1.161233 & 0.504527 & -3 & -1.540839  & 0.625280 & -3 & -1.909497 & 0.728140 \\
   -2 & 1.577787  & 0.689037 & -2 & 1.790982   & 0.728834 & -2 & 1.961880  & 0.748509 \\
   -1 & -2        & 0.879356 & -1 & -2         & 0.816066 & -1 & -2        & 0.763356 \\
    1 & 2         & 0.879356 & 1  & 2          & 0.816066 & 1  & 2         & 0.763356 \\
    2 & -1.577787 & 0.689037 & 2  & -1.790982  & 0.728834 & 2  & -1.961880 & 0.748509 \\
    3 & 1.161233  & 0.504527 & 3  & 1.540839   & 0.625280 & 3  & 1.909497  & 0.728140 \\
    4 & -0.939477 & 0.407302 & 4  & -1.388243  & 0.562493 & 4  & -1.873498 & 0.714152 \\
  \hline
\end{tabular}
\end{table}

\subsection{Other types of oscillatory chains}

Chains of the coupled Duffing oscillators are used for some physical
applications. For example in \cite{Sato}, the model of such a type
is exploited for describing discrete breathers in cantilever arrays.
Indeed, the authors of that paper consider an oscillatory chain with
on-site and inter-site forces both containing linear and cubic
terms. Now we want to demonstrate that the pair synchronization
method can be applied for the breather construction in the different
chains of such class of mechanical systems.

We will consider the following two models:
\begin{equation}\label{eq61}
\ddot{x}_i+x_i+x_i^3=\gamma [(x_{i+1}-x_i)^3-(x_{i}-x_{i-1})^3], \\
\end{equation}
and
\begin{equation}\label{eq62}
\ddot{x}_i+x_i^3=\gamma [(x_{i+1}-x_i)^3-(x_{i}-x_{i-1})^3], \\
\end{equation}
with periodic boundary conditions~(\ref{eq2}).

The model~(\ref{eq61}) contrary to (\ref{eq1}) represents the chain
of the Duffing oscillators with \emph{cubic} coupling. The
model~(\ref{eq62}) represents the so called $K_4$ chain, i.e. the
chain whose potential energy is a uniform function of the fourth
order. Discrete breathers in the $K_4$ chains with and without
on-site interactions were studied in many papers from different
points of view \cite{Kivshar, Flach-94, Aubry}. This chain is very
convenient for investigation because it allows separation of space
and time variables. In our paper \cite{Chechin}, the symmetry
discrete breathers were found for the model~(\ref{eq62}) with high
precision using the concept of the similar nonlinear normal modes
introduced by Rosenberg in \cite{Rozenberg}. The detailed
investigation of the breather stability is also presented in
\cite{Chechin}.

In Table~\ref{t40}, we present the velocity and amplitude profiles
for the chain~(\ref{eq61}) for some values of the coupling parameter
$\gamma$ in the chain with $N=7$ particles. It is worth comparing
this table with that for the linear coupled Duffing oscillators (see
Table~\ref{t2}). One can see that the replacing of the linear
coupling of the model~(\ref{eq1}) on the \emph{cubic} coupling of
the model~(\ref{eq61}) changes the degree of the breather spatial
localization.
\begin{table}
  \centering
  \caption{\label{t40} Symmetric discrete breathers for the chain (\ref{eq61})   for different values of the coupling parameter $\gamma$ ($N=7$).}
\begin{tabular}{|c|c|c|c|c|c|c|c|c|}
  \hline
  \multicolumn{3}{|c|}{$\gamma=0.3$; $\omega_b=1.86714$} & \multicolumn{3}{|c|}{$\gamma=0.5$; $\omega_b=1.89609$} & \multicolumn{3}{|c|}{$\gamma=0.7$; $\omega_b=2.22539$}\\
  \hline
  n & $\alpha_n$ & $A_n$ & n & $\alpha_n$ & $A_n$ & n & $\alpha_n$ & $A_n$ \\
  \hline
   -3 & $10^{-8}$        & $10^{-7}$         & -3 & -1.7$\times 10^{-7}$   & 9.3$\times 10^{-8}$    & -3& -4$\times 10^{-7}$     & 2$\times 10^{-7}$ \\
   -2 & 0.0071987 & 0.0043733  & -2 & 0.017014  & 0.0094272 & -2& 0.023577  & 0.012211 \\
   -1 & -0.598577 & 0.363639   & -1 & -0.776917 & 0.430465  & -1& -0.857977 & 0.444358 \\
    0 & 2         & 1.215014   & 0  & 2         & 1.108135  & 0 & 2         & 1.035827 \\
    1 & -0.598577 & 0.363639   & 1  & -0.776917 & 0.430465  & 1 & -0.857977 & 0.444358 \\
    2 & 0.0071987 & 0.0043733  & 2  & 0.017014  & 0.0094272 & 2 & 0.023577  & 0.012211 \\
    3 & $10^{-8}$        & $10^{-7}$         & 3  & -1.7$\times 10^{-7}$   & 9.3$\times 10^{-8}$   & 3 & -4$\times 10^{-7}$      & 2$\times 10^{-7}$ \\
  \hline
  \multicolumn{3}{|c|}{$\gamma=1$; $\omega_b=2.41195$} & \multicolumn{3}{|c|}{$\gamma=1.2$; $\omega_b=2.51505$} & \multicolumn{3}{|c|}{$\gamma=1.5$; $\omega_b=2.64864$} \\
  \hline
  n & $\alpha_n$ & $A_n$ & n & $\alpha_n$ & $A_n$ & n & $\alpha_n$ & $A_n$ \\
  \hline
    -3 & -8.9$\times 10^{-7}$   & 4.3$\times 10^{-7}$    & -3& -1.1$\times 10^{-6}$   & 5.2$\times 10^{-7}$    & -3 & -1.4$\times 10^{-6}$    & 6.3$\times 10^{-7}$\\
    -2 &  0.029453 & 0.014155  & -2& 0.031950  & 0.014764  & -2 & 0.034560   & 0.015210\\
    -1 & -0.917511 & 0.440940  & -1& -.940079  & 0.434406  & -1 & -0.962237  & 0.423472\\
     0 & 2         & 0.961164  & 0 & 2         & 0.924192  & 0  & 2          & 0.880183\\
     1 & -0.917511 & 0.440940  & 1 & -.940079  & 0.434406  & 1  & -0.962237  & 0.423472\\
     2 & 0.029453  & 0.014155  & 2 & 0.031950  & 0.014764  & 2  & 0.034560   & 0.015210\\
     3 & -8.9$\times 10^{-7}$   & 4.3$\times 10^{-7}$    & 3 & -1.1$\times 10^{-6}$   & 5.2$\times 10^{-7}$     & 3  & -1.4$\times 10^{-6}$    & 6.3$\times 10^{-7}$\\
  \hline
\end{tabular}
\end{table}

In Table~\ref{t41}, we present the similar information for the $K_4$
chain, i.e. for the chain with cubic force interactions and
\emph{without}
 the linear on-site forces. As was already mentioned, for the latter
 model there exist Rosenberg nonlinear normal modes. For these modes,
 the displacement of every particle of the chain
 at any instant $t$ is proportional to the displacement of an
 arbitrary chosen particle, say, the first particle. Therefore, for
 the dynamical regime described by a Rosenberg mode one can write
\begin{equation}\label{eq65}
x_ i(t)=k_i\cdot x_1(t), \ \ i=1..N,
\end{equation}
where $k_i$ are the constant coefficients which can be determined
from a certain systems of nonlinear algebraic equations (see
\cite{Chechin} for the further details).
\begin{table}
  \centering
  \caption{\label{t41} Symmetric discrete breathers for the chain (\ref{eq62})  for different values of the coupling parameter $\gamma$ ($N=5$).}
\begin{tabular}{|c|c|c|c|c|c|c|c|c|}
  \hline
  \multicolumn{3}{|c|}{$\gamma=0.3$; $\omega_b=1.86714$} & \multicolumn{3}{|c|}{$\gamma=0.5$; $\omega_b=1.97301$} & \multicolumn{3}{|c|}{$\gamma=0.7$; $\omega_b=2.13965$}\\
  \hline
  n & $\alpha_n$ & $A_n$ & n & $\alpha_n$ & $A_n$ & n & $\alpha_n$ & $A_n$ \\
  \hline
   -2 & 0.0071987 & 0.0043733  & -2 & 0.017014  & 0.010332 & -2& 0.023577  & 0.013202 \\
   -1 & -0.598577 & 0.363639   & -1 & -0.776917 & 0.471793 & -1& -0.857977 & 0.480441 \\
    0 & 2         & 1.215014   & 0  & 2         & 1.214528 & 0 & 2         & 1.119939 \\
    1 & -0.598577 & 0.363639   & 1  & -0.776917 & 0.471793 & 1 & -0.857977 & 0.480441 \\
    2 & 0.0071987 & 0.0043733  & 2  & 0.017014  & 0.010332 & 2 & 0.023577  & 0.013202 \\
  \hline
  \multicolumn{3}{|c|}{$\gamma=1$; $\omega_b=2.33466$} & \multicolumn{3}{|c|}{$\gamma=1.2$; $\omega_b=2.44172$} & \multicolumn{3}{|c|}{$\gamma=1.5$; $\omega_b=2.57983$} \\
  \hline
  n & $\alpha_n$ & $A_n$ & n & $\alpha_n$ & $A_n$ & n & $\alpha_n$ & $A_n$ \\
  \hline
    -2& 0.029452   & 0.015115  &-2 &  0.031948 & 0.981391  &  -2 & 0.034559   & 0.016050\\
    -1& -0.917511  & 0.470863  &-1 & -0.940078 & 0.461292   & -1 & -0.962237  & 0.446887\\
     0 & 2         & 1.026392  & 0 & 2         & 0.981391  &  0  & 2          & 0.928850\\
     1 & -0.917511 & 0.470863  & 1 & -0.940078 & 0.461292  & 1  & -0.962237   & 0.446887\\
     2 & 0.029452  & 0.015115  & 2 & 0.031948  & 0.981391  &  2  & 0.034559   & 0.016050\\
  \hline
\end{tabular}
\end{table}

From (\ref{eq65}), one can see that $\dot{x}_j(t)=k_i\dot{x}_1(t)$,
\ $i=1..N$ and, therefore, $\dot{x}_i(0)=k_i\dot{x}_1(0)$. In turn,
this means that the initial velocity profile is proportional to the
spatial profile of the Rosenberg mode which was found for some range
of $\gamma$ in \cite{Chechin}. The results of that paper and those
from Table~\ref{t41} of the present paper, obtained by the method of
pair synchronization, proved to be identical.

\section{A simple physical interpretation of the pair synchronization procedure\label{sec4}}

In Sec.~\ref{syn2}, we have described the method of pair
synchronization for discrete breather construction. At each step of
this method, we force two particles, who vibrate with
\emph{essentially different amplitudes}, to have \emph{exactly
identical frequencies}. Keeping in mind that in nonlinear regimes,
particles vibrating with different amplitudes possess, as a rule,
different frequencies, one can wonder what is the \emph{physical
nature} of the possibility of the above discussed pair
synchronization? The understanding of this nature leads us to the
explicit answer to the question formulated in the introduction of
the present paper: "How can discrete breathers exist as space
localized and, at the same time, strictly time-periodic dynamical
objects"?

Let us consider the chain~(\ref{eq1}) for $N=3$ and write down
explicitly the corresponding dynamical equations:
\begin{equation*}
\ddot{x}_1+x_1+x_1^3=\gamma (x_{2}-2x_1+x_{3}),
\end{equation*}
\begin{equation}\label{eq10}
\ddot{x}_2+x_2+x_2^3=\gamma (x_{3}-2x_2+x_{1}),
\end{equation}
\begin{equation*}
\ddot{x}_3+x_3+x_3^3=\gamma (x_{1}-2x_3+x_{2}).
\end{equation*}

These equations describe the vibrations of three Duffing oscillators
with linear coupling and parameter $\gamma$ determines the strength
of this coupling. The relation $x_2(t)=x_3(t)$ holds, since we
search for a symmetric breather. Therefore, tree
equations~(\ref{eq10}) reduce to the following two equations:
\begin{equation}\label{eq11a}
\ddot{x}_1+x_1+x_1^3=2\gamma (x_{2}-x_1),
\end{equation}
\begin{equation}\label{eq11b}
\ddot{x}_2+x_2+x_2^3=\gamma (x_{1}-x_2).
\end{equation}
We suppose
\begin{equation}\label{eq12}
0<\gamma\ll1, \ \ \ |x_2(t)|\ll|x_1(t)|.
\end{equation}
Actually, the last relation is the condition of strong space
localization of our three-particle breather. Then it is resonable to
neglect the term $x_2^3$ in Eq.~(\ref{eq11b}) and the term $2\gamma
x_2$ in Eq.~(\ref{eq11a}). As a consequence of such simplification,
Eqs.~(\ref{eq11a}-\ref{eq11b}) read \footnote{It is easy to prove
numerically that, for resonable values of the parameters $\gamma$
and $\alpha_1=\dot{x_1}(0)$, Eqs.~(\ref{eq12a}-\ref{eq12b}) allow
the breather solution, which is very close to that of the exact
equations~(\ref{eq10}).}
\begin{equation}\label{eq12a}
\ddot{x}_1+(1+2\gamma)x_1+x_1^3=0,
\end{equation}
\begin{equation}\label{eq12b}
\ddot{x}_2+(1+\gamma)x_2=\gamma x_1.
\end{equation}
Eq.~(\ref{eq12a}) represents the Duffing equation whose solution can
be expressed via Jacobi elliptic functions, while Eq.~(\ref{eq12b})
is the equation of the harmonic oscillator with \emph{periodic
external} force $\gamma\cdot x_1(t)$.

Let us make a next step of simplification replacing periodic
function $x_1(t)$ in (\ref{eq12b}) by its first Fourier harmonic or,
more precisely, let us assume
\begin{equation}\label{eq13}
x_1(t)=A\sin(\omega t)+B\cos(\omega t),
\end{equation}
where the correct frequency $\omega$ can be obtained as a result of
solving the Duffing equation~(\ref{eq12a}).

From the initial conditions
\begin{equation}\label{eq14}
x_1(0)=0, \ \ \ \dot{x_1}(0)=\alpha_1,
\end{equation}
we find for the solution~(\ref{eq13}) \ that
$A=\frac{\alpha_1}{\omega}$, \ $B=0$ and, therefore,
\begin{equation}\label{eq15}
x_1(t)=\frac{\alpha_1}{\omega}\sin(\omega t).
\end{equation}

\begin{figure}
\includegraphics[width=100mm,height=55mm]{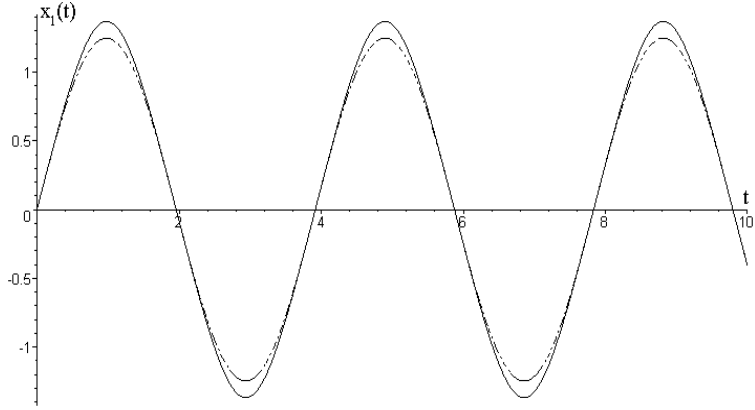}
\caption{\label{F40} The exact solution to the Duffing equation
(\ref{eq12a}) for $\gamma=0.1$, \ $\alpha_1=2$ (solid line) and its
approximation (\ref{eq15}) (dotted line).}
\end{figure}

In Fig.~\ref{F40} we depict the exact solution to Eq.~(\ref{eq12a})
(by a solid line) and the approximate solution~(\ref{eq15}) (by a
dotted line) for \ $\gamma=0.1$, \ $\alpha_1=2$. From this figure,
one can see how close can be to each other the above two solutions.

Substituting~(\ref{eq15}) into r.h.s. of Eq.~(\ref{eq12b}), we can
write
\begin{equation}\label{eq16}
\ddot{x_2}(t)+\omega_0^2\cdot
x_2=\frac{\gamma\alpha_1}{\omega}\sin(\omega t),
\end{equation}
where $\omega_0^2=1+\gamma$.  Eq.~(\ref{eq16}) is the equation of
the harmonic oscillator with fundamental frequency $\omega_0$ and
with periodic driven force proportional to $\sin(\omega t)$.

Taking into account the initial conditions $x_2(0)=0$, \
$\dot{x_2}(0)=\alpha_2$, we easily obtain the solution to
Eq.~(\ref{eq16}):
\begin{equation}\label{eq17}
x_2(t)=\biggl[\alpha_2-\frac{\gamma\alpha_1}{(\omega_0^2-\omega^2)}\biggr]\sin(\omega_0
t)+\frac{\gamma\alpha_1}{\omega(\omega_0^2-\omega^2)}\sin(\omega t).
\end{equation}
Here  the first term containing $\sin(\omega_0 t)$ represents a
solution to the equation~(\ref{eq16}) without right-hand side, i.e.
a solution of the \emph{free} harmonic oscillator with frequency
$\omega_0$, while the second term containing $\sin(\omega t)$
represents a particular solution to the full Eq.~(\ref{eq16}) with
the periodic driven force in its r.h.s.

Let us analyze the solution~(\ref{eq17}). In general case, the
frequencies $\omega_0$ and $\omega$ are \emph{incommensurable}.
Therefore, for existence of the discrete breathers as strictly
\emph{time-periodic} dynamical objects it is necessary to
\emph{eliminate} the contribution from the "vibrational
individuality" of the peripheral particles, i.e. the coefficient in
front of $\sin(\omega_0 t)$ must be equal to zero. From this
condition, we find the initial velocity $\alpha_2$ of the second
particle via that ($\alpha_1$) of the central particle:
\begin{equation}\label{eq21}
\alpha_2=-\frac{\gamma\alpha_1}{\omega^2-\omega_0^2}.
\end{equation}
Note that $\omega>\omega_0$ since the frequency of the hard Duffing
oscillator increases with increasing of the vibrational amplitude.
Therefore, $\alpha_2$ and $\alpha_1$ possess \emph{opposite signs}.

Certainly, Eq.~(\ref{eq21}) gives us only a certain approximation to
$\alpha_2$, because we take into account only one harmonic from the
Fourier decomposition of the function $x_1(t)$. Thus, our
approximate breather solution is
\begin{equation}\label{eq25a}
x_1(t)=\frac{\alpha_1}{\omega}\cdot\sin(\omega t),
\end{equation}
\begin{equation}\label{eq25b}
x_2(t)\equiv
x_3(t)=-\frac{\gamma\alpha_1}{\omega(\omega^2-\omega_0^2)}\cdot
\sin(\omega t).
\end{equation}

It is essential, that the vibrational amplitude of $x_2(t)$ is
$\gamma/(\omega^2-\omega_0^2)$ times less than the amplitude of the
central particle. In a certain sense, the relation $\gamma\ll1$
means the "space localization" of our breather in the three-particle
chain.

Actually, obtaining (\ref{eq25a}-\ref{eq25b}) we have applied the
synchronization procedure $S(1,2)$ (see Sec.~\ref{syn2}) in the
framework of our approximation to nonlinear dynamics of the Duffing
chain~(\ref{eq1}) with $N=3$. Considering the chain with $N=5$
particles (see Fig.~\ref{fig5}), we must also synchronize the
vibrations of the particles $2$ and $3$. Obviously, all arguments
which brought us to Eq.~(\ref{eq25a}-\ref{eq25b}) can be applied to
this synchronizing. Indeed, because of Eq.~(\ref{eq25b}), the second
particle $\Bigl[x_2(t)=x_5(t)\Bigr]$ already vibrates with the
frequency $\omega$ of the central particle. Therefore, the procedure
$S(2,3)$ reduces to the finding solution for $x_3(t)=x_4(t)$ by
solving the harmonic oscillator equation for $x_3(t)$ with
time-periodic force driving exerted by the function $x_2(t)$ from
Eq.~(\ref{eq25b}). Demanding \emph{full suppression} of the
fundamental frequency contribution of the third particle [function
$x_3(t)$] leads us to the relation similar to~(\ref{eq21})
\begin{equation}\label{eq29}
\alpha_3=-\frac{\gamma\alpha_2}{\omega^2-\tilde{\omega}_0^2},
\end{equation}
where $\tilde{\omega}_0^2=1+\gamma$.

Thus, the vibrational amplitudes of the third and fourth particles \
$\frac{\gamma}{\omega^2-\tilde{\omega}_0^2}$ \ times less than that
of the particles $2$ and $5$ (see Fig~\ref{fig5}).

Obviously, all this argumentation can be continued to the
consideration of the breathers in the  chains with arbitrary number
of particles. As a result, we conclude that vibrational amplitudes
of all the peripheral particles decrease, at least, by the factor
$\gamma\ll1$. In turn, this means that our discrete breather
possesses \emph{exponential decay} in vibrational amplitudes when we
pass on from the central particle to the more distant (peripheral)
particles.

One additional comment, concerning the fundamental frequencies
$\omega_0=\sqrt{1+2\gamma}$ and $\tilde{\omega_0}=\sqrt{1+\gamma}$
is appropriate at this point.

For the symmetric breather in the chain with $N=5$ particles (see
Fig.~\ref{fig5}), there hold relations $x_2(t)=x_5(t)$, \
$x_3(t)=x_4(t)$. Substituting these relations into the dynamical
equations (1), we obtain:
\begin{equation*}
\ddot{x}_1+x_1+x_1^3=2\gamma (x_{2}-x_1), \\
\end{equation*}
\begin{equation}\label{eq30}
\ddot{x}_2+x_2+x_2^3=\gamma (x_{3}-2x_2+x_{1}), \\
\end{equation}
\begin{equation*}
\ddot{x}_3+x_3+x_3^3=\gamma (x_{2}-x_3).
\end{equation*}
Then taking into account the relations $|x_2(t)|\ll|x_1(t)|$ \ and \
$|x_3(t)|\ll|x_2(t)|$, which are a consequence of the \emph{breather
localization}, one can reduce Eqs.~(\ref{eq30}) to the following
form:
\begin{equation*}
\ddot{x}_1+\omega_0^2x_1+x_1^3=0, \\
\end{equation*}
\begin{equation}\label{eq31}
\ddot{x}_2+\omega_0^2x_2=\gamma x_{1}, \\
\end{equation}
\begin{equation*}
\ddot{x}_3+\tilde{\omega}_0^2x_3=\gamma x_{2},
\end{equation*}
containing slightly different fundamental frequencies $\omega_0$ and
$\tilde{\omega}_0$.

\begin{figure}
\includegraphics{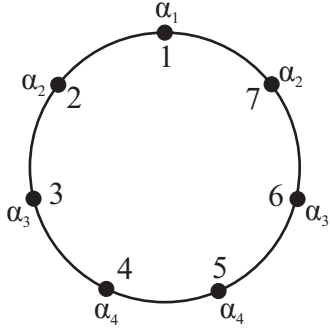}
\caption{\label{F50} Scheme of the initial conditions for the
seven-particle chain.}
\end{figure}

The similar procedure for the case $N=7$ with taking into
consideration the relations (see Fig.~\ref{F50}) $x_2(t)=x_7(t)$, \
$x_3(t)=x_6(t)$, \ $x_4(t)=x_5(t)$, brings us to the equations

\begin{equation}\label{eq32}
\begin{array}{c}
\ddot{x}_1+\omega_0^2x_1+x_1^3=0, \\
\ddot{x}_2+\omega_0^2x_2=\gamma x_{1}, \\
\ddot{x}_3+\omega_0^2x_3=\gamma x_{2},\\
\ddot{x}_4+\tilde{\omega}_0^2x_4=\gamma x_{3}.
\end{array}
\end{equation}

Proceeding in such a way for the chains with greater $N$, we
conclude that all approximate equations contain fundamental
frequency $\omega_0=\sqrt{1+2\gamma}$ excepting only the last
equation which contains the fundamental frequency
$\tilde{\omega}_0=\sqrt{1+\gamma}$. Bearing in mind that
$\gamma\ll1$, we obtain, from the equations similar to
Eqs.~(\ref{eq31}) and (\ref{eq32}), the above formulated conclusion
about exponential decay of the amplitudes of the peripheral
breather's particles.

Let us summurize the above results.  It has been manifested that the
existence of discrete breather is connected with the "suppressing of
the individuality" of all peripheral particles by vibrations of the
central particle of our breather. In other words, the central
particle forces upon the peripheral particles its own rhythm of
vibrations. To obtain the correct discrete breather one must tune
onto this rhythm with \emph{high precision} by suppressing all terms
in the solution to Eq.~(\ref{eq1}) corresponding to vibrations with
different frequencies of all the peripheral particles. Therefore, an
exact discrete breather seems to be a very unusual dynamical object
and, in practice, we deal with quasibreathers rather than with exact
breathers \cite{Chechin}. For the  former objects, it is typical
that all breathers particles vibrate with slightly different
frequencies and these frequencies slightly drift in time. In turn,
this means that in the spatially localized solution to
Eq.~(\ref{eq1}) there exist terms with slightly different
fundamental frequencies of the peripheral particles, whose
amplitudes are \emph{sufficiently small}.

We believe that the presence of small amplitude terms with the
frequencies $\omega_j$ different from the principal breather
frequency $\omega_b$,  peculiar to any quasibreather solutions,
cannot be the cause of the loss of their stability (see, for
example, Eq.~(\ref{eq17})). In other words, the loss of the strict
periodicity does not mean, in general, the loss of the spatial
localization of the considered dynamical objects. However, at the
present time, we cannot give a rigorous mathematical proof of this
proposition.

In conclusion, let us make two additional comments.

\begin{figure}
\includegraphics[width=120mm,height=55mm]{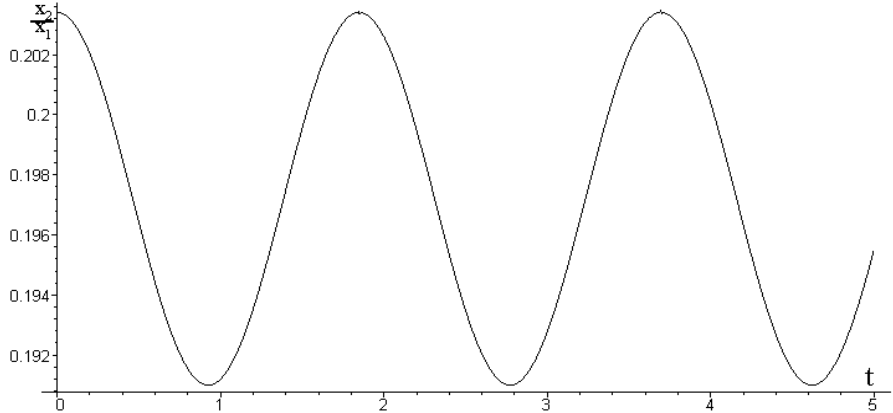}
\caption{\label{F51} The ratio $\frac{|x_2(t)|}{|x_1(t)|}$ for the
exact solution to Eqs.~(\ref{eq11a}, \ref{eq11b}) for
$\gamma=0.3$.}
\end{figure}

\begin{enumerate}
    \item For the chains with an \emph{uniform} potential, one can construct
    the Rosenberg nonlinear normal modes~\cite{Chechin}. Our
    chain~(\ref{eq1}) of the Duffing oscillators does not belong to
    such class of mechanical systems. Nevertheless, the approximate
    solutions obtained in this section [see, for example,
    Eqs.~(\ref{eq25a}-\ref{eq25b})] prove to be the Rosenberg mode,
    since ratios \ $\frac{x_j(t)}{x_1(t)}=k_j$ \ ($j=1..N$) \ do not
    depend on $t$. But if we take into account the next Fourier
    harmonics for $x_1(t)$, such ratios do \emph{depend on time}. We
    illustrate this fact in Fig.~\ref{F51} for the chain with $N=3$
    particles.
    \item In previous sections, we consider not only the
    single-frequency breathers, but many-frequencies breathers, as
    well. The existence of such dynamical objects can be also
    understood in the framework of our approximate methods for the
    discrete breathers construction discussed  in the present
    section. Indeed, let us return to the expression~(\ref{eq17})
    for the chain with $N=3$ particles. Now we will not annihilate
    the coefficient in front of $\sin(\omega_0 t)$. Instead of this, we choose such value of the initial velocity of the central
    particle that leads to the frequency $\omega$, which is a certain \emph{multiple} with respect to the
frequency $\omega_0$ of the
    peripheral  particle: $\omega=m\omega_0$ with $m$ being an
    integer number \footnote{It is possible owing to the specific dependence
    of $\omega$ on the vibrational amplitude for the Duffing oscillator (see
    Fig.~\ref{F10})}.
Then the resulting breather possesses the frequency $\omega_0$,
determined by the vibrations of two peripheral particles ($2$ and
$3$), while the frequency $\omega=m\omega_0$ of the central particle
turns out to be higher multiple of $\omega_0$. Thus, we obtain a
discrete breather whose particles vibrate with \emph{two different},
but divisible, frequencies ($\omega_0$ and $\omega=m\omega_0$), i.e.
we find a \emph{two-frequency} breather. In this case, the strictly
periodic discrete breather exists also only if we tune onto the
relation $\omega=m\omega_0$ with high precision, otherwise we obtain
a quasibreather with two incommensurable frequencies $\omega$ and
$\omega_0$.
\end{enumerate}

When this paper was finished, we became aware of the very
interesting paper \cite{Ovch-1969} by Ovchinnikov concerning the
localization of the vibrational energy in strongly excited molecules
and molecular crystals. The author has presented the elegant
arguments for the possibility of the above localization using as an
example a molecular dimer described by equations similar to
Eqs.~(\ref{eq11a}-\ref{eq11b}) of our paper. The everaging procedure
was applied to obtain an approximate analytical solution to these
dynamical equations in order to demonstrate the existence of the
periodic solution with different vibrational amplitudes of the dimer
particles. Above in this section, we have presented somewhat
different arguments for the same goal.

\section{Conclusion}
In this paper, we present a simple method for the discrete breather
construction which has been called the pair synchronization method
because of its transparent physical meaning. Indeed, it represents a
certain iterative procedure which, at each step, synchronizes
subsequently the motion of the pairs of particles involving in the
breather vibration. We believe, that this method can be applied for
construction exact breathers in the nonlinear lattices of different
types. The Duffing oscillatory chains with linear and nonlinear
coupling were considered for explanation how the pair
synchronization method can be used in practice.

A comparison of efficiency of the pair synchronization method with
other methods for the discrete breather construction based on the
Newton procedure for solving nonlinear algebraic equations, on the
steepest descent procedure etc. will be presented elsewhere.

With some additional approximation, the pair synchronization method
leads to a very simple physical interpretation of the possibility of
the exact breathers existence as strictly time-periodic and
spatially localized dynamical objects in nonlinear Hamiltonian
lattices. Indeed, constructing a strictly time-periodic breather, we
actually demand annihilation of the contributions with natural
vibrational frequencies  from all the peripheral particles to the
breather solution. This means the existence of some kind of
dictatorship of the central particle (in general, of the breather
core) which must "suppress any individuality" of the peripheral
particles by above mentioned annihilation. In other words, all terms
of the exact breather solution with frequencies different from that
of the central particle (which appear from the natural frequencies
of the peripheral particles) must be turn into zero in  compulsory
order). The breather core compels the peripheral particles vibrate
with its own frequency. In Sec.~\ref{sec4}, we demonstrate, for the
case of weakly coupled oscillatory chains, that this compulsion
leads to the exponential decay of the vibrational amplitudes of the
peripheral breather particles.

On the other hand, if we do not tune the initial conditions on the
\emph{exact} breather solution, i.e. if the vibrational
contributions from the peripheral particles are not equal to zero,
we obtain a certain \emph{quasibreather} \cite{Chechin}. This
spatially localized dynamical object is characterized by different
frequencies appearing from the peripheral particles. The difference
in frequencies is brought about by the phenomenon typical for
nonlinear dynamics, namely, by the dependence of the frequency on
the vibrational amplitude.

In \cite{Chechin}, we proposed to characterize the proximity of the
quasibreather to an exact time-periodic breather with the aid of the
mean square deviations in frequencies of the individual particles
and that of the fixed particle in time (there exist a certain
temporal drift of the frequency of every chosen particle). We also
presented there some arguments (at least, for the case of $K_4$
chain) for the possibility of the quasibreather stability despite of
the absence of the time periodicity. However, we cannot now present
any refine mathematical proof on quasibreather stability for
infinite time.

\begin{figure}
\includegraphics[width=140mm,height=55mm]{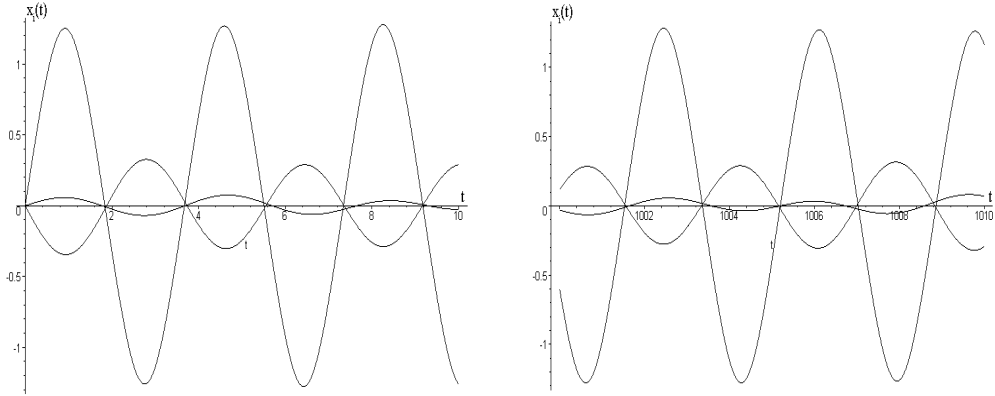}
\caption{\label{F52} An extremely long-lived quasibreather for the
Duffing chain~(\ref{eq1}) with $N=5$ particles which, possibly, is a
stable dynamical object [$\gamma=0.3$, \ $\alpha_1=2$, \
$\alpha_2=-0.5760$, $\alpha_3=0.1$].}
\end{figure}

On the other hand, one often reveals that the lifetime of a
quasibreather, if it actually decays in time, can be exponantionally
large for small deviations from the exact breather. For example, in
Fig.~\ref{F52}, we depict the time evolution of the quasibreather in
the weakly coupled Duffing chain which lives, at least, up to
$t\sim10^6$.

Finally, since we cannot tune exactly on the strictly time periodic
breather solution in any physical experiment (and even in numerical
experiments), quasibreathers seem to be much more relevant spatially
localized dynamical objects than the exact discrete breathers. Some
additional detail for this point of view can be found in
\cite{Chechin}.

\begin{acknowledgments}
We are very grateful to Prof.~V.~P.~Sakhnenko for his friendly
support and to S.~Flach for very useful critical comments and for
his hospitality during our visit to Dresden in December 2006. The
authors thank strongly N.A. Zemlyanova for her valuable linguistic
consultations. G.S.D. is very grateful to Russian foundation
"Dynasty" for the financial support. This work has been supported
by Southern Federal University (Russia) under the project
K-07-T-25.
\end{acknowledgments}

\end{document}